\documentclass[3p,twocolumn]{elsarticle}
\usepackage{hhline}
\usepackage{fancyvrb}
\newfont{\myBbb}{msbm10 scaled 1200}
\newcommand{\mod}[1]{\ (\bmod\ #1)}
\newcommand{\Z}{{\mbox{\myBbb Z}}}

\newcounter{bla}

\journal{Computer Physics Communications}

\begin{document}

\begin{frontmatter}

\title{RNGSSELIB: Program library for random number generation, SSE2 realization}

\author{L.Yu. Barash}
\ead{barash@itp.ac.ru}
\author{L.N. Shchur}
\address{Landau Institute for Theoretical Physics, 142432
Chernogolovka, Russia}

\begin{abstract}

The library RNGSSELIB for random number generators (RNGs) based upon the SSE2 command set 
is presented.
The library contains realization of a number of modern and most reliable generators.
Usage of SSE2 command set allows to substantially improve performance of the generators.
Three new RNG realizations are also constructed.
We present detailed analysis of the speed depending on compiler usage and associated
optimization level, as well as results of extensive statistical testing
for all generators using available test packages. 
Fast SSE implementations produce exactly the same output sequence as the original algorithms.
\end{abstract}

\end{frontmatter}

{\bf PROGRAM SUMMARY}

\begin{small}
\noindent
{\em Manuscript Title:} RNGSSELIB: Program library for random number generation, SSE2 realization \\
{\em Authors:} L.Yu. Barash, L.N. Shchur \\
{\em Program Title:} RNGSSELIB \\
{\em Journal Reference:}                                      \\
{\em Catalogue identifier:}                                   \\
{\em Licensing provisions:}                                   \\
{\em Programming language:} C                                \\
{\em Computer:} PC \\
{\em Operating system:} UNIX, Windows                                      \\
{\em RAM:} 1 Mbytes   \\
{\em Number of processors used:} 1 \\
{\em Supplementary material:}                                 \\
{\em Keywords:} Statistical methods, Monte Carlo, 
  Random numbers, Pseudorandom numbers, Random number generation, Streaming SIMD Extensions \\
{\em Classification:} 4.13 Statistical Methods   \\
{\em External routines/libraries:}                            \\
{\em Subprograms used:}                                       \\
{\em Nature of problem:}  Any calculation requiring uniform pseudorandom 
number generator, in particular, Monte Carlo calculations. \\
   \\
{\em Solution method:}
The library contains realization of a number of modern and reliable generators:
\verb#mt19937#, \verb#mrg32k3a# and \verb#lfsr113#.
Also new realizations for the method based on parallel evolution of an ensemble
of dynamical systems are constructed: \verb#GM19#, \verb#GM31# and \verb#GM61#.
The library contains both usual realizations and realizations based on SSE command set.
Usage of SSE commands allows to substantially improve performance of all generators.
\\
   \\
{\em Restrictions:} For SSE realizations of the generators,
Intel or AMD CPU supporting SSE2 command set is required.
In order to use the realization \verb#lfsr113sse#, CPU must
support SSE4 command set. \\
   \\
{\em Unusual features:}\\
   \\
{\em Additional comments:}\\
   \\
{\em Running time:} Running time is of the order of $20$ sec
for generating $10^9$ pseudorandom numbers with a PC based 
on Intel Core i7-940 CPU.
Running time is analysed in detail in Sec.~5 of the paper.\\
   \\

\end{small}

\section{Introduction}

Pseudo random numbers, generated recursively by deterministic rules,
represent one of important ingredients of numerical simulations
widely used in physics and material science~\cite{cm}.
Design of the random number generators (RNG)
producing pseudo random numbers which approximate ``true
randomness''~\cite{Knuth} is a great challenge for the computational
physics and computer science in general.

High speed and statistical robustness are the most important
characteristics needed for good RNG. Also, long period length of the
generated sequences is among the most desired features.

In the paper we present a library of modern and most reliable RNGs known today.
Namely, the realizations of the modern generators MT19937, MRG32k3a, LFSR113,
GM19SSE, GM31SSE and GM61SSE are presented. MT19937 is the 2002 version of
the Mersenne Twister generator of Matsumoto and Nishimura~\cite{MT}, which is
based on the recent generalizations to the GFSR method.
MRG32k3a is the combined multiple recursive generator
proposed in~\cite{CombinedLCG}, and LFSR113 is the
combined Tausworthe generator of L'Ecuyer~\cite{LFSR113}.
Each of the generators GM19SSE, GM31SSE and GM61SSE
is based on an ensemble of sequences generated by multiple recursive method.
The method is closely related to the method of random number generation
based on evolution of the ensemble of the chaotic dynamical systems~\cite{BarashShchur}.
We introduce three RNG realizations based on the method: GM19SSE, GM31SSE and GM61SSE.

An important feature of our library is the ability to speed up
the RNG generation using Streaming SIMD Extensions 2 (SSE2)
technology, introduced in Intel Pentium 4 processors in 2001~\cite{Pentium4}.
AMD added SSE2 support to their processors with the introduction of their
Opteron and Athlon64 ranges of 64-bit CPUs in 2003~\cite{AMD64}.
All of Intel and AMD processors, fabricated later than 2003, support SSE2 instruction set.
SSE2 technology allows to use 128-bit XMM-registers effectively and to accelerate computations.
A similar technique was previously used for implementing some RNGs~\cite{RS,BarashShchur}.
In the present work we demonstrate that this approach can essentially speed up most RNGs.

Program codes for the generators and proper initializations
can be found in~\cite{AlgSite}. We would like to stress that our SSE implementations of algorithms produce 
exactly the same output sequence  as the original algorithms. 
We would appreciate any comments on user experiences.

\section{Classical and modern random number generators}

The most widely used RNGs can be divided into two classes.
The first class is represented by linear congruential generator (LCG),
and the second by shift register (SR), or Tausworthe generator.

Linear Congruential Generators (LCGs) are the best-known and (still)
most widely available RNGs in use today. An example of the
realization of an LCG generator is the UNIX {\tt rand} generator
$y_n=(1103515245\; y_{n-1}+12345)\mod{2^{31}}$.
The practical recommendation is that LCGs should be avoided
for applications dealing with the geometric behavior of random
vectors in high dimensions because of the bad geometric structure of
the vectors that they produce~\cite{Knuth,Coveyou}.
Another problem is that classical LCGs have relatively small period lengths.
Also, the lower-order bits of the generated sequence have a far shorter
period than the sequence as a whole if the modulo is set to a power of $2$.
An immediate generalization of LCG is Multiple Recursive Generator (MRG).

Generalized Feedback Shift Register (GFSR) sequences are widely
used in many areas of computational and simulational physics~\cite{Kirk,Golomb}. 
These RNGs are quite fast and possess huge periods given a proper choice
of the underlying primitive trinomial. This makes them
particularly well suited for applications that require many
pseudorandom numbers. But several flaws have been observed
in the statistical properties of these generators, which can result in
systematic errors in Monte Carlo simulations. Typical examples include
the Wolff single cluster algorithm simulations for the 2D Ising model~\cite{SR1}
and 3D Ising model~\cite{S99}, random and self-avoiding walks~\cite{Grass,RandomWalksTest}, and the 3D
Blume--Capel model using local Metropolis updating~\cite{SR3}.

Modern RNGs are modifications and generalizations to the LCG and GFSR methods,
they have much better periodic and statistical properties~\cite{LEcuyerMain}.
In this work we discuss properties, efficient realizations and statistical tests for some of them.

\subsection{Mersenne Twister random number generator MT19937}
\label{MT19937Sec}

Algorithm MT19937 is a Mersenne Twister (MT) generator by Matsumoto and Nishimura~\cite{MT}.
In a sense, MT algorithm represents a modified and twisted GFSR generator.
MT generates the vectors of word size by the recurrence
\begin{equation}
\mathbf{x}_{k+n} := \mathbf{x}_{k+m} \oplus (\mathbf{x}_k^u|\mathbf{x}_{k+1}^l)A.
\label{MTrecc}
\end{equation}
For MT19937 values of parameters are chosen as follows:
$n=624$, $m=397$, $(\mathbf{x}_k^u|\mathbf{x}_{k+1}^l)$ is the 32-bit integer
obtained by concatenating one upper bit of $\mathbf{x}_k$ and 31 lower bits of
$(\mathbf{x}_{k+1}^l)$, and $\oplus$ is exclusive or operation. $A$ is a companion matrix:
$\mathbf{x}A=\mbox{shiftright}(\mathbf{x})$ if the least significant bit of $\mathbf{x}$ is $0$,
otherwise $\mathbf{x}A=\mbox{shiftright}(\mathbf{x}) \oplus \mathbf{a}$, where
vector $\mathbf{a}$ consists of 32 bits and represents 32-bit integer constant $2567483615$.
MT19937 is a generator with a very long period $2^{19937}-1$,
and it provides 623-dimensional equidistribution of pseudorandom numbers up to 32-bit accuracy.
As for most generators, Mersenne Twister is sensitive to poor initialization,
so proper initialization is very important.

\subsection{Combined Multiple Recursive Generator MRG32k3a}
\label{Mrg32k3aSec}

Algorithm MRG32k3a represents a Combined Multiple Recursive Generator (CMRG)
found in~\cite{CombinedLCG} by means of extensive search of parameters for
such generator.
MRG32k3a algorithm combines two MRGs (see right column
in Table~\ref{MRGTable}). The state of the first MRG
is described by variables \verb#x0#, \verb#x1#, \verb#x2#,
whereas the variables \verb#y0#, \verb#y1#, \verb#y2#
represent the state of the second MRG. At each step,
the MRG states are shifted as shown in the right column of Table~\ref{MRGTable},
where the calculated values of \verb#p1# and \verb#p2# are the new outputs of the MRGs.

\subsection{Combined Tausworthe generator LFSR113}
\label{LFSR113Sec}

Another modern generator is a combined Tausworthe generator, i.e.,
a combined GFSR sequence. In~\cite{LFSR113} an extensive search for good parameters
for such generator is carried out, and, as a result, an algorithm LFSR113 for
such generator is presented. Right column in Table~\ref{LFSRTable} describes the algorithm.
The state of the generator is described by variables \verb#z1#,\verb#z2#,\verb#z3#,\verb#z4#.

\subsection{Generators GM19, GM31, and GM61 based on the ensemble of MRGs}
\label{GMSec}

Finally, we include in the library improved versions of
the generators GM19 and GM31 worked out in~\cite{BarashShchur},
and also the new realization GM61.
It is suggested in~\cite{BarashShchur} to construct
RNGs based on an ensemble of hyperbolic automorphisms
of the unit two-dimensional torus,
\begin{equation}
{x_i^{(n)}\choose y_i^{(n)}}=M
{x_i^{(n-1)}\choose y_i^{(n-1)}}\mod{g},
\label{MatrixRecurr}
\end{equation}
where $i=0,1,\dots,(s-1)$, the elements of two-dimensional matrix $M$ are integers, $\det M=1$,
and $|\mbox{Tr} M|>2$.
The output of the generator is defined as
\begin{equation}
a^{(n)}=\sum_{i=0}^{s-1} \lfloor 2x_i^{(n)}/g\rfloor \cdot 2^i.
\label{OutputFunction}
\end{equation}
Dynamical system~(\ref{MatrixRecurr})
is widely known under name cat map (there are two reasons for this terminology: first,
CAT is an acronym for Continuous Automorphism of the Torus; second, the
chaotic behavior of these maps is traditionally described by showing the
result of their action on the face of the cat~\cite{Arnold}).
The main properties of the generator based on cat maps can be predicted theoretically
because there is close relation between the periodic properties of cat maps and
arithmetical properties of algebraic numbers~\cite{BarashShchur}.
The basic recurrence~(\ref{MatrixRecurr}), representing an evolution of a simple nonlinear
dynamical system on a discrete $g\times g$ lattice, is equivalent to linear recurrence modulo $g$
with the same characteristic polynomial $f(x)=x^2-kx+q$ as that of the matrix $M$:
\begin{eqnarray}
x^{(n)}&=&kx^{(n-1)}-qx^{(n-2)}\mod{g},
\label{Recurrence}
\\
y^{(n)}&=&ky^{(n-1)}-qy^{(n-2)} \mod{g},
\end{eqnarray}
where $k=\mbox{Tr} M$, $q=\det M=1$~\cite{Grothe,Lecuyer90,BarashShchur}.
Therefore, in a sense, the method represents
a modified and generalized MRG, where only part of the information of a generator
state influences the generator output.

\begin{table}[ht]
\caption{Parameters of the generators GM19, GM31 and GM61.}
\label{GMParams}
\begin{tabular}{|l|c|c|c|c|}
\hline
Generator &  $k$   &  $q$  &   $g$      &      Period         \\
\hline
GM19      &  $15$  & $28$  & $2^{19}-1$ & $2.7\cdot 10^{11}$  \\
GM31      &  $11$  & $14$  & $2^{31}-1$ & $4.6\cdot 10^{18}$  \\
GM61      &  $24$  & $74$  & $2^{61}-1$ & $5.3\cdot 10^{36}$  \\
\hline
\end{tabular}
\end{table}

The modified RNG based on the same idea, where lattice
parameter $g$ is chosen to be Mersenne prime, and
the characteristic polynomial $f(x)=x^2-kx+q$ is chosen
to be primitive over $\Z_g$, possess very good statistical
and periodic properties. In this case
the basic recurrence is closely related
to so-called matrix generator of pseudorandom numbers
studied in~\cite{Knuth,Grothe,NiederrBook}.
Primitivity of the characteristic polynomial guarantees maximal
possible period $g^2-1$ of the output sequence, also it is
required for the primitivity of $f(x)$ that $q\ne 1$.
Table~\ref{GMParams} shows parameter values chosen
for generators GM19, GM31 and GM61.
For generating random numbers it is convenient
to employ the linear recurrence (\ref{Recurrence}).
As above, the output function is defined
with (\ref{OutputFunction}), i.e. each bit
of the output corresponds to its own recurrence,
and $s=32$ recurrences are calculated in parallel.

There is an easy algorithm to calculate $x^{(n)}$ in (\ref{Recurrence})
very quickly from $x^{(0)}$ and $x^{(1)}$ for any large $n$.
Indeed, if $x^{(2n)}=k_n x^{(n)}-q_n x^{(0)}\mod{g}$, then
$x^{(4n)}=(k_n^2-2q_n) x^{(2n)}-q_n^2 x^{(0)}\mod{g}$.
As was mentioned already in~\cite{BarashShchur}, this helps to
initialize the generator. To initialize all $32$ recurrences,
the following initial conditions are used:
$x_i^{(0)}=x^{(iA)}$, $x_i^{(1)}=x^{(iA+1)}, i=0,1,\dots,31$.
Here $A$ is a value of the order of $(p^2{-}1)/32$, which is
not very close to a divisor of $p^2-1$ or to a large power of $2$.

\section{SSE-realizations of RNGs}
\label{DescrSec}

\newsavebox{\FirstAlgBox}
\newsavebox{\SecondAlgBox}
\newsavebox{\ThirdAlgBox}
\newsavebox{\FourthAlgBox}
\newsavebox{\Example}

Modern Intel and AMD processors support SSE2 instruction set. SSE2 is acronym for Streaming SIMD (Single Instruction Multiple Data) Extensions 2.
SSE2 realization of instruction set allows user to perform the same operation in parallel on four 32-bit numbers placed in 128-bit XMM registers, or on two 64-bit numbers~\cite{Pentium4}. It is known that proper programming of 32-bit RNG using SSE command set may essentially decrease computation time needed for generating random number.  For example, in paper~\cite{RS} SSE realization of the combined RNG based on two Shift Register generators with shift register lengths 9689 and 4423 accelerated random number generation by factor 3.5.

In this section we describe realization of random number generators discussed in the previous section using SSE2 command set.
These effective realizations are equivalent to classical algorithms discussed
in previous section, i.e. all output values are the same.
For example, MT19937SSE algorithm produces the same output values as well-known MT19937
introduced in~\cite{MT},
in contrast to other Mersenne Twister SSE algorithms introduced in~\cite{SIMD-MT}.

Program codes for all realizations and
examples of practical implementations, including proper initialization,
can be found in~\cite{AlgSite}.

\begin{lrbox}{\FirstAlgBox}
\begin{minipage}{8cm}
\begin{verbatim}

   asm("movaps  2500(%0),%%xmm0\n" \
       "movaps  4(%0),%%xmm1\n" \
       "pand    upper_mask,%%xmm0\n" \
       "pand    lower_mask,%%xmm1\n" \
       "por     %%xmm0,%%xmm1\n" \
       "movaps  %%xmm1,%%xmm0\n" \
       "psrld   $1,%%xmm1\n" \
       "pand    and1_mask,%%xmm0\n" \
       "pcmpgtd zero_mask,%%xmm0\n" \
       "pand    andA_mask,%%xmm0\n" \
       "pxor    %%xmm1,%%xmm0\n" \
       "movaps  1588(%0),%%xmm1\n" \
       "cmpl    $227,%2\n" \
       "jl      MyL1%=\n" \
       "movaps  -908(%0),%%xmm1\n" \
       "MyL1%=: pxor    %%xmm1,%%xmm0\n" \
       "movups  %%xmm0,(%0)\n" \
       "movaps  %%xmm0,2500(%0)\n" \
       ""::"r"(mt+i),"r"(consts),"r"(i));

\end{verbatim}
\end{minipage}
\end{lrbox}

\begin{lrbox}{\SecondAlgBox}
\begin{minipage}{7cm}
\begin{verbatim}

\end{verbatim}
\end{minipage}
\end{lrbox}

\begin{table}[ht]
\caption{SSE2 realization for the main part of algorithm MT19937SSE
(tempering is omitted in the table).}
\label{MTTable}
\begin{tabular}{|c|}
\hline
\usebox{\FirstAlgBox} \\
\hline
\end{tabular}
\end{table}

\subsection{SSE2 based MT19937}

In Table~\ref{MTTable}
we present main ideas of our version for MT19937 algorithm (see Sec.\ref{MT19937Sec}),
based on SSE2 command set.
The whole array of 624 doubleword integers is divided into fours of integers,
and four recurrent relations (\ref{MTrecc}) are calculated in parallel.

\subsection{SSE2 based MRG32k3a}

Right column in Table~\ref{MRGTable} shows usual MRG32k3a algorithm (see Sec.~\ref{Mrg32k3aSec}).
The recurrent relations are:
$x_{i+3}=ax_i+bx_{i+1}\mod{m_1}$ and $y_{i+3}=cy_i+dy_{i+2}\mod{m_2}$,
where $a=-810728$, $b=1403580$, $c=-1370589$, $d=527612$.
Therefore,
$x_{i+4}= ax_1 + bx_2 \mod{m_1}$ and
$x_{i+6}= \alpha_0 x_0 + \alpha_1 x_1 + \alpha_2 x_2 \mod{m_1}$,
$y_{i+4}= \beta_0 y_0 + \beta_1 y_1 + \beta_2 y_2 \mod{m2}$,
$y_{i+6}= \gamma_0 y_0 + \gamma_1 y_1 + \gamma_2 y_2 \mod{m2}$,
where constants $\alpha_0$, $\alpha_1$, $\alpha_2$, $\beta_0$, $\beta_1$, $\beta_2$,
$\gamma_0$, $\gamma_1$ and $\gamma_2$ can be easily calculated.
This allows to calculate $x_4$, $x_5$, $x_6$ and $x_7$ from $x_0,x_1,x_2$,
and also $y_4$, $y_5$, $y_6$ and $y_7$ from $y_0,y_1,y_2$.
Calculation of fours of numbers was found to be very effective.
Main parts of the SSE-realization of the MRG32k3a algorithm
are shown in the left column in Table~\ref{MRGTable}.

Since there are no SSE commands for taking modulo in XMM registers, we use the relation
$z \equiv (f\cdot N+g) \mod{m}$ for $z=f\cdot 2^{32}+g$ and $N=2^{32}-m$.
We calculate $z_1$, $z_2$, $z_3$ using formulas
$z_0=f_0\cdot 2^{32}+g_0$, $z_1=f_0\cdot N + g_0=f_1\cdot 2^{32}+g_1$,
$z_2=f_1\cdot N + g_1=f_2\cdot 2^{32}+g_2$,
if $z_2<m$ then $z_3=z_2$ else $z_3=z_2-m$. This results in our case
in $z_3<m$. SSE commands admit simultaneous calculation of $\verb#x4#\ (\bmod\ \verb#m1#)$
and $\verb#x5#\ (\bmod\ \verb#m2#)$. Finally the RNG output is calculated as \verb#p1 - p2 + m1#
for $\verb#p1#\le \verb#p2#$ and \verb#p1 - p2# for $\verb#p1#>\verb#p2#$.

\subsection{SSE4 based LFSR113}

Our realizations of LFSR113 and LFSR113SSE (see Sec.\ref{LFSR113Sec})
are presented in Table~\ref{LFSRTable}.
The SSE realization utilizes \verb#pmulld# and \verb#pblendw# processor
instructions of SSE4 command set. Therefore, the realization LFSR113SSE requires CPU supporting
SSE4, for example, a processor with Intel Core microarchitecture.
The realization works {\em only} for Intel processors and is not
valid for AMD processors.
For shifting four integers of an XMM register
to the left by different numbers of bits, we use the SSE4 instruction \verb#pmulld#.
For shifting to the right by different numbers of bits, we use
SSE4 command \verb#pblendw#.
The details of the algorithms are shown in Table~\ref{LFSRTable}.

\subsection{SSE2 based GM19, GM31 and GM61}

Table~\ref{GMTable} illustrates the key ideas for speeding up algorithms GM19, GM31
and GM61 (see Sec.~\ref{GMSec}) using SSE2 command set.
Actions of the fast SSE2 algorithms shown in the left
column are equivalent to actions of the slow algorithms shown in the right column.

\begin{lrbox}{\FirstAlgBox}
\begin{minipage}{7.8cm}
\begin{verbatim}

const long long add1=9007203111403311U;
const long long add2=9007202867859652U;

// here m1 divides add1, m2 divides add2

const long long m1=   4294967087U;
const long long m2=   4294944443U;
const long long a =  -810728;
const long long b =  1403580;
const long long c = -1370589;
const long long d =   527612;
long long x0, x1, x2, y0, y1, y2;

unsigned mrg32k3a (){

  long k; long long p1, p2;

  p1 = (add1 + b*x1 + a*x0) % m1;

  x0 = x1; x1 = x2; x2 = p1;

  p2 = (add2 + d*y2 + c*y0) % m2;

  y0 = y1; y1 = y2; y2 = p2;

  if (p1 <= p2) return (p1 - p2 + m1);
     else return (p1 - p2);

}
\end{verbatim}
\end{minipage}
\end{lrbox}

\begin{lrbox}{\SecondAlgBox}
\begin{minipage}{9.7cm}
\begin{Verbatim}[fontsize=\small]

unsigned s[16] __attribute__ ((aligned(16)));
// { x0,0,x1,0, x2,0,x3,0, y0,0,y1,0, y2,0,y3,0 }

unsigned consts[72] __attribute__ ((aligned(16))) =
{ x4ap,0,x4ap,0,  x4bm,0,x4bm,0,
  x4cp,0,x4cp,0,  x4addl,x4addh,x4addl,x4addh,
  x6ap,0,x6ap,0,  x6bp,0,x6bp,0,
  x6cm,0,x6cm,0,  x6addl,x6addh,x6addl,x6addh,
  m1,0,m1,0, [...]
};

// Calculating x4 and x5 from x1 and x2:

  asm("movups  8(%0),%%xmm1\n"\
      "movaps  16(%0),%%xmm2\n"\
      "pmuludq 16(%1),%%xmm1\n"\
      "pmuludq 32(%1),%%xmm2\n"\
      "paddq   48(%1),%%xmm2\n"\
      "psubq   %%xmm1,%%xmm2\n"\
      "movaps %%xmm2,%%xmm3\n"\
      "psrlq $32,%%xmm3\n"\
      "pmuludq 128(%1),%%xmm3\n"\
      "psubq %%xmm3,%%xmm2\n"\
      "movaps %%xmm2,%%xmm3\n"\
      "psrlq $32,%%xmm3\n"\
      "pmuludq 128(%1),%%xmm3\n"\
      "psubq %%xmm3,%%xmm2\n"\
      "psubq 128(%1),%%xmm2\n"\
      "movaps %%xmm2,%%xmm3\n"\
      "psrad $31,%%xmm3\n"\
      "psrlq $32,%%xmm3\n"\
      "pand  128(%1),%%xmm3\n"\
      "paddq %%xmm3,%%xmm2\n"\
      "movaps  %%xmm2,(%0)\n"\
      ""::"r"(s),"r"(consts));

// Calculating (x0-y0) (mod m1):

  asm("movaps (%0),%%xmm1\n"\
      "psubq  32(%0),%%xmm1\n"\
      "movaps %%xmm1,%%xmm4\n"\
      "psrad $31,%%xmm4\n"\
      "psrlq $32,%%xmm4\n"\
      "pand  128(%1),%%xmm4\n"\
      "paddq %%xmm4,%%xmm1\n"\
      ""::"r"(s),"r"(consts));

\end{Verbatim}
\end{minipage}
\end{lrbox}

\begin{table*}[p]
\caption{Realizations MRG32K3aSSE (left column) and MRG32k3a (right column) producing equivalent outputs.}
\label{MRGTable}
\begin{tabular}{|c|c|}
\hline
main parts of SSE2-based MRG32k3aSSE & MRG32k3a-L with x86-64 command set \\ \hline
\usebox{\SecondAlgBox} & \usebox{\FirstAlgBox}\\
\hline
\end{tabular}
\end{table*}

\begin{lrbox}{\FirstAlgBox}
\begin{minipage}{8cm}
\begin{verbatim}

unsigned z[4] __attribute__ ((aligned(16)))
= {12345,12345,12345,12345};

unsigned lfsr113(){
  unsigned output;
  asm("movaps (%1),%%xmm1\n" \
      "movaps (%2),%%xmm2\n" \
      "movaps (%4),%%xmm0\n" \
      "pand   %%xmm1,%%xmm2\n" \
      "pmulld (%3),%%xmm2\n" \
      "pmulld %%xmm1,%%xmm0\n" \
      "pxor   %%xmm0,%%xmm1\n" \
      "psrld  $12,%%xmm1\n" \
      "pblendw $192,%%xmm1,%%xmm3\n" \
      "psrld  $1,%%xmm1\n" \
      "pblendw $3,%%xmm1,%%xmm3\n" \
      "psrld  $8,%%xmm1\n" \
      "pblendw $48,%%xmm1,%%xmm3\n" \
      "psrld  $6,%%xmm1\n" \
      "pblendw $12,%%xmm1,%%xmm3\n" \
      "pxor   %%xmm2,%%xmm3\n" \
      "movaps %%xmm3,(%1)\n" \
      "pshufd $255,%%xmm3,%%xmm0\n" \
      "pshufd $170,%%xmm3,%%xmm1\n" \
      "pshufd $85,%%xmm3,%%xmm2\n" \
      "pxor %%xmm0,%%xmm3\n" \
      "pxor %%xmm1,%%xmm2\n" \
      "pxor %%xmm2,%%xmm3\n" \
      "pextrd $0,%%xmm3,%0\n" \
      "":"=&r"(output):"r"(z),"r"(a),
      "r"(b),"r"(c));
  return output;
}

\end{verbatim}
\end{minipage}
\end{lrbox}

\begin{lrbox}{\SecondAlgBox}
\end{lrbox}

\begin{lrbox}{\ThirdAlgBox}
\begin{minipage}{8cm}
\begin{verbatim}

unsigned z1=12345,z2=12345,
         z3=12345,z4=12345;

unsigned lfsr113()
{
   unsigned long b;

   b = ((z1 <<  6) ^ z1) >> 13;
   z1 = ((z1 & 4294967294U) << 18) ^ b;
   b = ((z2 <<  2) ^ z2) >> 27;
   z2 = ((z2 & 4294967288U) <<  2) ^ b;
   b = ((z3 << 13) ^ z3) >> 21;
   z3 = ((z3 & 4294967280U) <<  7) ^ b;
   b = ((z4 <<  3) ^ z4) >> 12;
   z4 = ((z4 & 4294967168U) << 13) ^ b;
   return (z1 ^ z2 ^ z3 ^ z4);
}

\end{verbatim}
\end{minipage}
\end{lrbox}

\begin{table*}[p]
\caption{Algorithms LFSR113 and LFSR113SSE}
\label{LFSRTable}
\begin{tabular}{|l|l|}
\hline
SSE4 based LFSR113SSE & LFSR113 based on usual instruction set \\
\hline
\usebox{\FirstAlgBox} & \usebox{\ThirdAlgBox}\\
\hline
\end{tabular}
\end{table*}

\begin{table*}[p]
\caption{Equivalent realizations for several algorithms
for processor with SSE2 support (left column) and
in ANSI C language (right column).
First row presents calculating four matrix multiplications.
Second row presents the packing $16$ high bits of $16$ integers
into one integer.
These or similar equivalences are used in constructing the SSE2 algorithms
for GM19, GM31 and GM61 algorithms~\cite{AlgSite}.}

\begin{lrbox}{\FirstAlgBox}
\begin{minipage}{7cm}
\begin{verbatim}

  unsigned long x[4],y[4];

  [.......]

  asm("movaps (%0),%%xmm0\n" \
      "movaps (%1),%%xmm1\n" \
      "paddd  %%xmm1,%%xmm0\n" \
      "paddd  %%xmm1,%%xmm0\n" \
      "movaps %%xmm0,%%xmm2\n" \
      "pslld  $2,%%xmm0\n" \
      "paddd  %%xmm1,%%xmm0\n" \
      "movaps %%xmm0,(%0)\n" \
      "psubd  %%xmm2,%%xmm0\n" \
      "movaps %%xmm0,(%1)\n" \
      ""::"r"(x),"r"(y));

\end{verbatim}
\end{minipage}
\end{lrbox}

\begin{lrbox}{\SecondAlgBox}
\begin{minipage}{7cm}
\begin{verbatim}

  unsigned long i,newx[4],x[4],y[4];

  [.......]

  for(i=0;i<4;i++){
    newx[i]=4*x[i]+9*y[i];
    y[i]=3*x[i]+7*y[i];
    x[i]=newx[i];
  }

\end{verbatim}
\end{minipage}
\end{lrbox}

\begin{lrbox}{\ThirdAlgBox}
\begin{minipage}{7cm}
\begin{verbatim}

unsigned long x[16],output;

  [.......]

  asm("movaps (%1),%%xmm0\n" \
      "movaps 16(%1),%%xmm1\n" \
      "movaps 32(%1),%%xmm2\n" \
      "movaps 48(%1),%%xmm3\n" \
      "psrld  $31,%%xmm0\n" \
      "psrld  $31,%%xmm1\n" \
      "psrld  $31,%%xmm2\n" \
      "psrld  $31,%%xmm3\n" \
      "packssdw %%xmm1,%%xmm0\n" \
      "packssdw %%xmm3,%%xmm2\n" \
      "packsswb %%xmm2,%%xmm0\n" \
      "psllw  $7,%%xmm0\n" \
      "pmovmskb %%xmm0,%0\n" \
      "":"=r"(output):"r"(x));

\end{verbatim}
\end{minipage}
\end{lrbox}

\begin{lrbox}{\FourthAlgBox}
\begin{minipage}{7cm}
\begin{verbatim}

const unsigned long halfg=2147483648;
unsigned long x[16],i,output=0,bit=1;

  [.......]

for(i=0;i<16;i++){
  output+=((x[i]<halfg)?0:bit;
  bit*=2;
}

\end{verbatim}
\end{minipage}
\end{lrbox}

\label{GMTable}
\begin{tabular}{|c|c|}
\hline Four matrix multiplications in parallel, SSE2 version&
Same with usual instruction set \\ \hline
\usebox{\FirstAlgBox} & \usebox{\SecondAlgBox}\\
\hline Packing 16 high bits into one integer, SSE2 version &
Same with usual instruction set\\
\hline
\usebox{\ThirdAlgBox} & \usebox{\FourthAlgBox}\\
\hline
\end{tabular}
\label{EquivAlgs}
\end{table*}

\section{How to use the library. Function call interface.}

\subsection{Realization MRG32K3A-SSE}

Table~\ref{ExampleTable} shows the example of using the MRG32K3A-SSE realization 
in ANSI C language.
Table~\ref{genRand} shows header file names for the generators.
Using the realization requires including 
the header file \verb#mrg32k3asse.h# in the code.

\begin{table}[tb]
\caption{Example of using the realization MRG32K3A-SSE.}
\label{ExampleTable}
\begin{tabular}{|c|}
\hline
\begin{minipage}{8cm}
\begin{Verbatim}[fontsize=\small]

#include<stdio.h>
#include"mrg32k3asse.h"

int main(){
  unsigned i; mrg32k3asse_state state;
  mrg32k3asse_SetState(&state);
  mrg32k3asse_genInit(123,123,123,123,123,123); 
  mrg32k3asse_genPrintState();
  for(i=0;i<99999999;i++) mrg32k3asse_genRand();
  printf("10^8 Numbers Generated\n");
  printf("Next Output Value: %u\n",
         mrg32k3asse_genRand());
  mrg32k3asse_genPrintState();
  return 0;
}

\end{Verbatim}
\end{minipage}\\
\hline
\end{tabular}
\end{table}

The function \verb#mrg32k3asse_SetState(...)# in Table~\ref{ExampleTable}
saves the pointer to the structure \verb#state#
in order that the other functions
use it as a pointer to the structure that keeps 
the generator state. Such function should be executed prior to 
using any other functions that manipulate with a generator state.
It can also be used in order to switch between using several RNGs
of the same type. 
Table~\ref{SetState} describes the function call interface.

The function
\verb#mrg32k3asse_genInit(...)# in Table~\ref{ExampleTable} initializes the generator state.
Table~\ref{genInit} shows details for the initialization function.
The values of \verb#x0#, \verb#x1#, \verb#x2#, \verb#y0#, \verb#y1#,
\verb#y2# described in Sec.~\ref{Mrg32k3aSec} and Table~\ref{MRGTable} are required
to initialize any version of the generator MRG32K3A.

The function 
\verb#mrg32k3asse_genRand()# in Table~\ref{ExampleTable} 
generates a pseudorandom 32-bit integer
and makes respective transformation to the generator state.
Detailed interface for this function is presented in Table~\ref{genRand}.

The function \verb#mrg32k3asse_genRand4(#\verb#unsigned*# \verb#arr)#
generates block of four pseudorandom 32-bit integers and puts them in 
\verb#arr[0]#, \verb#arr[1]#, \verb#arr[2]# and \verb#arr[3]#.
The function \verb#mrg32k3asse_genRand32(unsigned* arr)#
generates block of 32 pseudorandom 32-bit integers and puts them in 
\verb#arr[0]#, \verb#arr[1]#,  \verb#...#, \verb#arr[31]#.
Detailed interface for the functions is presented in Table~\ref{genRand}.
Generation of blocks of numbers is further discussed in Sec.~\ref{SpeedSec}
and Tables~\ref{mt19937MKLTable} and~\ref{mrg32k3aMKLTable}.

The function 
\verb#mrg32k3asse_genPrintState()# prints out the generator state.
The function call interface is shown in Table~\ref{SetState}.

\subsection{Realization MRG32K3A}

The similar example as in Table~\ref{ExampleTable} 
for the realization MRG32K3A, which is exact
reproduction of the algorithm published in~\cite{CombinedLCG} and
is based on usual command set,
can be found in the file \verb#mrg32k3a.c# in~\cite{AlgSite}.

The header file for the realization is \verb#mrg32k3a.h#,
as it is shown in Table~\ref{genRand}.

Table~\ref{SetState} describes the function call interface for 
the function \verb#mrg32k3a_SetState(...)#
that saves a pointer to the generator state.

Table~\ref{genInit} shows details for the initialization functions
\verb#mrg32k3a_genInit(...)#.
The values of \verb#x0#, \verb#x1#, \verb#x2#, \verb#y0#, \verb#y1#,
\verb#y2# described in Sec.~\ref{Mrg32k3aSec} and Table~\ref{MRGTable} are required
to initialize the generator MRG32K3A.

The function \verb#mrg32k3a_genRand()# generates a pseudorandom
floating point number between $0$ and $1$ and makes
respective transformation to the generator state.
The function call interface is presented in Table~\ref{genRand}.

The function \verb#mrg32k3a_genPrintState()# prints out the generator state.
The function call interface is shown in Table~\ref{SetState}.

\subsection{Realizations MT19937 and MT19937-SSE}

The similar examples as in Table~\ref{ExampleTable} 
for MT19937 and MT19937-SSE can be found
in files \verb#mt19937.c# and \verb#mt19937sse.c# in~\cite{AlgSite}.

The header files for the realizations are \verb#mt19937.h# and \verb#mt19937sse.h#,
as it is shown in Table~\ref{genRand}.

Table~\ref{SetState} describes the function call interfaces for 
the functions \verb#mt19937_SetState(...)# and \verb#mt19937sse_SetState(...)#
that save a pointer to the corresponding generator state.

Table~\ref{genInit} shows details for the initialization functions
\verb#mt19937_genInit(...)# and \verb#mt19937sse_genInit(...)#.
Versions of the \verb#MT19937# generator are initialized 
with four unsigned integer values~\cite{MTInit}.

The functions 
\verb#mt19937_genRand()#  and \verb#mt19937sse_genRand()#
generate a pseudorandom 32-bit integer
and make respective transformation to the corresponding generator state.
Detailed interface for the functions is presented in Table~\ref{genRand}.

The function \verb#mt19937sse_genRand624()# generates block of 624
pseudorandom 32-bit unsigned integers and puts them in \verb#state->out[0]#,
\verb#state->out[1]#, \verb#...#, \verb#state->out[623]#,
where \verb#state->out# is array of unsigned integers, and 
the pointer \verb#state# to the structure of the type \verb#mt19937sse_state#
should have previously been saved with the function \verb#mt19937sse_SetState#.
Generation of blocks of numbers is further discussed in Sec.~\ref{SpeedSec}
and Tables~\ref{mt19937MKLTable} and~\ref{mrg32k3aMKLTable}.

The functions 
\verb#mt19937_genPrintState()# and \verb#mt19937sse_genPrintState()#
print out the corresponding generator state.
Detailed interface for the functions is shown in Table~\ref{SetState}.

\subsection{Realizations LFSR113 and LFSR113-SSE}

The similar examples as in Table~\ref{ExampleTable} 
for LFSR113 and LFSR113-SSE can be found
in files \verb#lfsr113.c# and \verb#lfsr113sse.c# in~\cite{AlgSite}.

The header files for the realizations are \verb#lfsr113.h# and \verb#lfsr113sse.h#,
as it is shown in Table~\ref{genRand}.

Table~\ref{SetState} describes the function call interfaces for 
the functions \verb#lfsr113_SetState(...)# and \verb#lfsr113sse_SetState(...)#
that save a pointer to the corresponding generator state.

Table~\ref{genInit} shows details for the initialization functions
\verb#lfsr113_genInit(...)# and \verb#lfsr113sse_genInit(...)#.
Initializing any version of the generator \verb#LFSR113#
requires initial values of \verb#z1#, \verb#z2#, \verb#z3#, \verb#z4#
described in Sec.~\ref{LFSR113Sec} and Table~\ref{LFSRTable} 
for the \verb#LFSR113# algorithm.

The functions 
\verb#lfsr113_genRand()#  and \verb#lfsr113sse_genRand()#
generate a pseudorandom 32-bit integer
and make respective transformation to the corresponding generator state.
Detailed interface for the functions is presented in Table~\ref{genRand}.

The functions 
\verb#lfsr113_genPrintState()# and \verb#lfsr113sse_genPrintState()#
print out the corresponding generator state.
Detailed interface for the functions is shown in Table~\ref{SetState}.

\subsection{Realizations GM19, GM19-SSE, GM31, GM31-SSE, GM61 and GM61-SSE}

The similar examples as in Table~\ref{ExampleTable} 
for GM19, GM19-SSE, GM31, GM31-SSE, GM61 and GM61-SSE can be found in files
\verb#gm19.c#, \verb#gm19sse.c#, \verb#gm31.c#, \verb#gm31sse.c#, 
\verb#gm61.c# and \verb#gm61sse.c# in~\cite{AlgSite}.

The header files for the realizations are
\verb#gm19.h#, \verb#gm19sse.h#, \verb#gm31.h#, \verb#gm31sse.h#, 
\verb#gm61.h# and \verb#gm61sse.h#, as it is shown in Table~\ref{genRand}.

Table~\ref{SetState} describes the function call interfaces for 
the functions \verb#gm19_SetState(...)#, \verb#gm19sse_SetState(...)#,
\verb#gm31_SetState(...)#, \verb#gm31sse_SetState(...)#,
\verb#gm61_SetState(...)# and \verb#gm61sse_SetState(...)#
that save a pointer to the corresponding generator state.

Table~\ref{genInit} shows details for the initialization functions
\verb#gm19_genInit(...)#, \verb#gm19sse_genInit(...)#,
\verb#gm31_genInit(...)#, \verb#gm31sse_genInit(...)#,
\verb#gm61_genInit(...)# and \verb#gm61sse_genInit(...)#.
For generators \verb#GM19# and \verb#GM19SSE# it is required
that $0\le \verb#seed# \le 9$, for \verb#GM31# and \verb#GM31SSE#
it is required that $0\le \verb#seed# \le 99$, for \verb#GM61# and \verb#GM61SSE# --
$0\le \verb#seed# \le 999$.

The functions 
\verb#gm19_genRand()#, \verb#gm19sse_genRand()#,
\verb#gm31_genRand()#, \verb#gm31sse_genRand()#,
\verb#gm61_genRand()# and \verb#gm61sse_genRand()#
generate a pseudorandom 32-bit integer
and make respective transformation to the corresponding generator state.
Detailed interface for the functions is presented in Table~\ref{genRand}.

The functions 
\verb#gm19_genPrintState()#, \verb#gm19sse_genPrintState()#,
\verb#gm31_genPrintState()#, \verb#gm31sse_genPrintState()#,
\verb#gm61_genPrintState()# and \verb#gm61sse_genPrintState()#
print out the corresponding generator state.
Detailed interface for the functions is shown in Table~\ref{SetState}.

\begin{table*}[p]
\caption{Function call interfaces for generating 32-bit pseudorandom numbers.}
\label{genRand}
\hspace{-1cm}
\begin{tabular}{|l|l|c|c|}
\hline
\small
Generator name & File to include & State type & Generate 32-bit random number(s) \\
\hline
\verb#GM19# & \verb#gm19.h# & \verb#gm19_state#  & \verb#unsigned gm19_genRand();# \\
\verb#GM19 (SSE)# & \verb#gm19sse.h# & \verb#gm19sse_state# & \verb#unsigned gm19sse_genRand();#\\
\verb#GM31# & \verb#gm31.h# & \verb#gm31_state#  & \verb#unsigned gm31_genRand();# \\
\verb#GM31 (SSE)# & \verb#gm31sse.h#  & \verb#gm31sse_state#  & \verb#unsigned gm31sse_genRand();# \\
\verb#GM61# & \verb#gm61.h#  & \verb#gm61_state# & \verb#unsigned gm61_genRand();# \\
\verb#GM61 (SSE)# & \verb#gm61sse.h#  & \verb#gm61sse_state#  & \verb#unsigned gm61sse_genRand();# \\
\verb#LSFR113# & \verb#lfsr113.h#  & \verb#lfsr113_state#  & \verb#unsigned lfsr113_genRand();# \\
\verb#LFSR113 (SSE)# & \verb#lfsr113sse.h# & \verb#lfsr113sse_state#  & \verb#unsigned lfsr113sse_genRand();# \\
\verb#MRG32K3A# & \verb#mrg32k3a.h# & \verb#mrg32k3a_state#  &  \verb#double mrg32k3a_genRand();# \\
\verb#MRG32K3A (SSE)# & \verb#mrg32k3asse.h# & \verb#mrg32k3asse_state#  & \verb#unsigned mrg32k3asse_genRand();# \\ 
\verb#MRG32K3A (SSE,4)# & \verb#mrg32k3asse.h# & \verb#mrg32k3asse_state# &  \verb#void mrg32k3asse_genRand4(unsigned* arr);# \\ 
\verb#MRG32K3A (SSE,32)# & \verb#mrg32k3asse32.h# & \verb#mrg32k3asse32_state# & \verb#void mrg32k3asse32_genRand32(unsigned* arr);# \\
\verb#MT19937# & \verb#mt19937.h# & \verb#mt19937_state# & \verb#unsigned long mt19937_genRand();# \\
\verb#MT19937 (SSE)# & \verb#mt19937sse.h# & \verb#mt19937sse_state# & \verb#unsigned mt19937sse_genRand();# \\
\verb#MT19937 (SSE,624)# & \verb#mt19937sse.h# & \verb#mt19937sse_state# & \verb#void mt19937sse_genRand624();# \\
\hline
\end{tabular}
\end{table*}

\begin{table*}[p]
\caption{Function call interfaces for keeping the pointer to a state and for printing out
a generator state.}
\label{SetState}
\begin{tabular}{|c|c|}
\hline
\small
Set state & Print out state \\
\hline
\verb#void gm19_SetState(gm19_state* state);# & \verb#void gm19_genPrintState();# \\
\verb#void gm19sse_SetState(gm19sse_state* state);# & \verb#void gm19sse_genPrintState();# \\
\verb#void gm31_SetState(gm31_state* state);# & \verb#void gm31_genPrintState();# \\
\verb#void gm31sse_SetState(gm31sse_state* state);# & \verb#void gm31sse_genPrintState();# \\
\verb#void gm61_SetState(gm61_state* state);# & \verb#void gm61_genPrintState();# \\
\verb#void gm61sse_SetState(gm61sse_state* state);# & \verb#void gm61sse_genPrintState();#\\
\verb#void lfsr113_SetState(lfsr113_state* state);# & \verb#void lfsr113_genPrintState();# \\
\verb#void lfsr113sse_SetState(lfsr113sse_state* state);# & \verb#void lfsr113sse_genPrintState();#\\
\verb#void mrg32k3a_SetState(mrg32k3a_state* state);# & \verb#void mrg32k3a_genPrintState();#\\
\verb#void mrg32k3asse_SetState(mrg32k3asse_state* state);# & \verb#void mrg32k3asse_genPrintState();#\\
\verb#void mrg32k3asse32_SetState(mrg32k3asse32_state* state);# & \verb#void mrg32k3asse32_genPrintState();#\\
\verb#void mt19937_SetState(mt19937_state* state);# & \verb#void mt19937_genPrintState();# \\
\verb#void mt19937sse_SetState(mt19937sse_state* state);# & \verb#void mt19937sse_genPrintState();# \\
\hline
\end{tabular}
\end{table*}

\begin{table*}[p]
\caption{Function call interfaces for initializing the generators.}
\label{genInit}
\begin{tabular}{|l|l|}
\hline
State type & Initialize state  \\
\hline
\verb#gm19_state# &  \verb#void gm19_genInit(unsigned seed);# \\
\verb#gm19sse_state# & \verb#void gm19sse_genInit(unsigned seed);# \\
\verb#gm31_state# &  \verb#void gm31_genInit(unsigned seed);# \\
\verb#gm31sse_state# & \verb#void gm31sse_genInit(unsigned seed);# \\
\verb#gm61_state# & \verb#void gm61_genInit(unsigned seed);# \\
\verb#gm61sse_state# & \verb#void gm61sse_genInit(unsigned seed);# \\
\verb#lfsr113_state# & \verb#void lfsr113_genInit(unsigned z1,unsigned z2,unsigned z3,unsigned z4);# \\
\verb#lfsr113sse_state# & \verb#void lfsr113sse_genInit(unsigned z1,unsigned z2,unsigned z3,unsigned z4);# \\
\verb#mrg32k3a_state# & \verb#void mrg32k3a_genInit(double x0,double x1,double x2,#\\
&\hspace{4.1cm}\verb#double y0,double y1,double y2);# \\
\verb#mrg32k3asse_state# & \verb#void mrg32k3asse_genInit(unsigned x0,unsigned x1,unsigned x2,#\\
&\hspace{4.6cm}\verb#unsigned y0,unsigned y1,unsigned y2);# \\
\verb#mrg32k3asse32_state# & \verb#void mrg32k3asse32_genInit(unsigned x0,unsigned x1,unsigned x2,#\\
&\hspace{5.0cm}\verb#unsigned y0,unsigned y1,unsigned y2);# \\
\verb#mt19937_state# & \verb#void mt19937_genInit(unsigned long init0, unsigned long init1,#\\
&\hspace{3.9cm}\verb#unsigned long init2, unsigned long init3);# \\
\verb#mt19937sse_state# & \verb#void mt19937sse_genInit(unsigned long init0, unsigned long init1,#\\
&\hspace{4.5cm}\verb#unsigned long init2, unsigned long init3);# \\
\hline
\end{tabular}
\end{table*}

\section{Speed and properties}
\label{SpeedSec}

CPU times needed for
generating $10^9$ random numbers are shown in Tables~\ref{Tspeed1},~\ref{Tspeed2}~and~\ref{Tspeed3}
for Intel Core 2 Duo E7400, Intel Core i7-940 and AMD Turion X2 RM-70 processors respectively.
The results are presented for different compilers
and optimization options for all RNGs of our library.
We use GNU C compiler gcc version 4.3.3,
and Intel C compiler icc version 11.0.
In all cases we observe decrease of computation time for the modified algorithms
as compared with original algorithms. In particular, using GNU C compiler
\verb+gcc+ with optimization level \verb#-O0#
we find that MRG32K3ASSE is about 4.5 times faster than MRG32K3A, whereas
LFSR113SSE is about $40\%$ faster than LFSR113
and MT19937SSE is about 3 times faster than MT19937.
Therefore, the modified generators turn out to be very fast.
Algorithms GM19SSE, GM31SSE and GM61SSE
are 10--22 times more efficient than corresponding algorithms based
on usual command set, and become competitive with other modern generators.

Table~\ref{mt19937MKLTable} contains CPU-times for generating $10^9$
random numbers via blocks of numbers,
both for our SSE2-based algorithm MT19937SSE and the MT19937 generator
from Intel MKL library. The output values of the generator realizations
are identical.

Table~\ref{mrg32k3aMKLTable} contains CPU-times for generating $10^9$
random numbers via blocks of numbers,
both for our SSE2-based algorithm MRG32K3ASSE and the MRG32K3A generator
from Intel MKL library. The output values of the generator realizations
are identical.

\begin{table*}[p]
\caption{CPU time (sec) for generating $10^9$ random numbers. Processor:~Intel Core~2~Duo~E7400.
Compilers: gcc 4.3.3, icc 11.0.}
\vspace{0.2cm}
\label{Tspeed1}
\begin{tabular}{|l|c|c|c|c||c|c|c|c||}
\hline
                 & gcc -O0 & gcc -O1 & gcc -O2 & gcc -O3 & icc -O0 & icc -O1 & icc -O2 & icc -O3\\
\hline
MRG32k3a         &  47.4   &  35.7   &  36.0   &  27.2   &  55.9   &   32.6  &  26.0   & 26.0   \\
MRG32k3aSSE      &  10.5   &  8.8    &  7.2    &  7.2    &   10.4  &   8.7   &  7.4    &  7.7   \\
MRG32k3a-4SSE    &  7.3    &  6.9    &  6.9    &  6.9    &   7.3   &   6.9   &  6.9    &  6.9   \\
LFSR113          &  10.9   &  4.6    &  4.8    &  3.5    &  10.8   &   5.1   &  4.6    &  4.6   \\
LFSR113SSE       &   8.0   &   7.4   &  7.3    &  7.3    &  8.3    &   7.7   &  7.5    &  7.5   \\
MT19937          &  16.6   &   6.1   &   6.1   &  2.7    &  15.8   &   6.8   &   2.9   &  2.8   \\
MT19937SSE       &   5.8   &   5.2   &   6.1   &  2.3    &  5.9    &   5.3   &   2.3   &  2.3   \\
GM19             & 616.9   &  242.2  & 200.1   & 144.4   &  628.9  & 226.9   & 125.1   & 124.9  \\
GM19SSE          &  32.8   &   30.6  &  27.8   &  27.7   &   33.5  &  29.8   &  28.5   &  28.2  \\
GM31             & 1067.9  &  255.6  & 204.9   & 151.4   &  1110.5 & 318.0   & 175.8   & 183.0  \\
GM31SSE          &  49.9   &  45.5   &  41.8   &  41.8   &   49.5  &  44.4   &  43.2   &  45.0  \\
GM61             & 1794.0  &  987.5  & 923.6   & 844.5   &  1908.9 & 1002.1  & 853.4   & 864.4  \\
GM61SSE          & 184.5   &  183.0  & 181.6   & 181.6   &  186.0  & 192.3   & 181.8   & 188.7  \\
\hline
\end{tabular}
\end{table*}

\begin{table*}[p]
\caption{CPU time (sec) for generating $10^9$ random numbers.
Processor:~Intel Core~i7-940. Compilers: gcc 4.3.3, icc 11.0.}
\vspace{0.2cm}
\label{Tspeed2}
\begin{tabular}{|l|c|c|c|c||c|c|c|c||}
\hline
                 & gcc -O0 & gcc -O1 & gcc -O2 & gcc -O3 & icc -O0 & icc -O1 & icc -O2 & icc -O3\\
\hline
MRG32k3a         &  47.9   &  36.3   &  35.3   &  25.0   &  56.1   &  33.1   &  22.8   &  28.1  \\
MRG32k3aSSE      &  9.1    &  7.4    &  5.8    &   5.8   &   8.8   &   7.4   &   6.0   &  5.9   \\
MRG32k3a-4SSE    &  6.2    &  5.8    &  5.8    &   5.7   &   6.4   &   5.9   &   5.9   &  5.7   \\
LFSR113          &  10.4   &  4.8    &  6.8    &   3.1   &  10.2   &   5.0   &   4.6   &  4.5   \\
LFSR113SSE       &   8.0   &  6.8    &  6.8    &   6.9   &   7.3   &   6.9   &   6.6   &  6.5   \\
MT19937          &  13.7   &  5.7    &  6.9    &   2.6   &  17.5   &   6.5   &   2.9   &  2.9   \\
MT19937SSE       &   5.2   &  4.8    &  5.5    &   2.0   &   4.9   &   4.7   &   2.4   &  2.0   \\
GM19             &  578.5  &  210.1  &  163.1  &  117.1  &  604.3  &  259.4  &  110.7  & 123.3  \\
GM19SSE          &  32.3   &  28.7   &  26.7   &  29.9   &  33.6   &  33.0   &  27.8   &  34.1  \\
GM31             &  999.0  &  244.6  &  181.6  &  134.3  &  978.9  &  298.8  &  143.4  & 151.0  \\
GM31SSE          &  39.1   &  36.4   &  38.5   &  58.7   &  47.1   &  36.4   &  35.4   &  35.9  \\
GM61             &  1599.7 &  893.8  &  836.9  &  766.6  & 1606.5  &  870.8  &  770.1  & 795.1  \\
GM61SSE          &  120.6  &  123.0  &  116.8  &  117.2  &  124.6  &  120.2  &  130.7  & 117.7  \\
\hline
\end{tabular}
\end{table*}

\begin{table*}[p]
\caption{CPU time (sec) for generating $10^9$ random numbers. Processor:~AMD Turion X2 RM-70.
Compilers: gcc 4.3.3, icc 11.0.}
\vspace{0.2cm}
\label{Tspeed3}
\begin{tabular}{|l|c|c|c|c||c|c|c|c||}
\hline
                 & gcc -O0 & gcc -O1 & gcc -O2 & gcc -O3 & icc -O0 & icc -O1 & icc -O2 & icc -O3\\
\hline
MRG32k3a         &  89.0   &  60.9   &  60.9   &  47.0   &  89.1   &  69.2   &  41.5   &  41.6  \\
MRG32k3aSSE      &  25.9   &  22.3   &  18.4   &  18.3   &  25.6   &  22.3   &  19.0   &  19.0  \\
MRG32k3a-4SSE    &  18.3   &  18.2   &  18.2   &  18.2   &  18.1   &  18.2   &  18.1   &  18.3  \\
LFSR113          &  14.6   &   8.7   &   9.6   &   5.3   &  14.9   &   9.1   &   6.9   &   6.8  \\
MT19937          &  31.0   &  17.8   &  10.8   &   7.1   &  31.0   &  18.7   &   5.2   &   4.9  \\
MT19937SSE       &  11.3   &  10.3   &  11.1   &   6.6   &  10.8   &   9.9   &   6.0   &   6.0  \\
GM19             & 1651.5  & 385.1   & 259.9   &  222.1  & 1759.9  &  375.4  &  198.6  & 198.4  \\
GM19SSE          &   96.0  & 105.3   &  87.5   &  87.5   &  103.5  &  102.8  &   88.6  &  88.9  \\
GM31             & 1512.5  &  436.7  & 313.2   &  237.3  & 1573.8  &  506.0  &  228.7  & 243.9  \\
GM31SSE          &  136.9  &  130.0  & 130.0   &  131.1  &  137.1  &  130.9  &  128.3  & 128.5  \\
GM61             & 5138.9  & 3826.0  & 3991.9  & 3567.9  & 5221.6  & 3849.2  & 3604.7  & 3586.9 \\
GM61SSE          &  418.3  &  413.6  & 411.1   &  409.7  &  427.5  &  416.6  &  407.7  & 410.2  \\
\hline
\end{tabular}
\end{table*}

\begin{table*}
\caption{Comparing CPU-times for different versions of MT19937
for generating $10^9$ random numbers via blocks of $N$ numbers.
Speed is also compared
with the respective realization of MT19937 generator in Intel MKL library.
Processor: Intel Core i7-940.}
\label{mt19937MKLTable}
\begin{center}
\begin{tabular}{|l|c|c|c|c||c|c|c|c||}
\hline
N & \multicolumn{4}{c||}{Intel MKL library} &  \multicolumn{4}{c||}{Our version}\\
\cline{2-9}
      & icc -O0 & icc -O1 & icc -O2 & icc -O3 & icc -O0 & icc -O1 & icc -O2 & icc -O3 \\
\hline
1     &  24.7   &  22.3   &  24.0   &  22.1   &  4.9    &  4.7    &  2.4    &   2.0    \\
624   &  1.8    &   1.8   &  1.8    &  1.8    &  1.8    &  1.6    &  1.4    &   1.4    \\
\hline
\end{tabular}
\end{center}
\end{table*}

\begin{table*}
\caption{Comparing CPU-times for different versions of MRG32K3A
for generating $10^9$ random numbers via blocks of $N$ numbers.
Speed is also compared
with the respective realization of MRG32K3A generator in Intel MKL library.
Processors: Intel Xeon 5160 (upper table), Intel Core i7-940 (lower table).}
\label{mrg32k3aMKLTable}
\begin{center}
\begin{tabular}{|l|c|c|c|c||c|c|c|c||}
\hline
N & \multicolumn{4}{c||}{Intel MKL library} &  \multicolumn{4}{c||}{Our version}\\
\cline{2-9}
      & icc -O0 & icc -O1 & icc -O2 & icc -O3 & icc -O0 & icc -O1 & icc -O2 & icc -O3 \\
\hline
1     &  37.5   &  33.9   &  33.0   &  33.0   &  10.0   &  8.2    &  7.2    &   7.2    \\
4     &  18.4   &  18.4   &  17.8   &  17.8   &  7.3    &  6.6    &  6.6    &   6.6    \\
32    &  10.6   &  10.6   &  10.6   &  10.6   &  6.6    &  6.6    &  6.5    &   6.5    \\
1000  &  6.4    &  6.6    &  6.6    &  6.6    &  6.6    &  6.6    &  6.6    &   6.6    \\
\hline
\end{tabular}
\begin{tabular}{|l|c|c|c|c||c|c|c|c||}
\hline
N & \multicolumn{4}{c||}{Intel MKL library} &  \multicolumn{4}{c||}{Our version}\\
\cline{2-9}
      & icc -O0 & icc -O1 & icc -O2 & icc -O3 & icc -O0 & icc -O1 & icc -O2 & icc -O3 \\
\hline
1     &  31.2   &  27.7   &  28.1   &  29.6   &  8.8    &  7.4    &  6.0    &   5.9    \\
4     &  16.3   &  16.0   &  15.6   &  15.7   &  6.4    &  5.9    &  5.9    &   5.7    \\
32    &  8.6    &   8.6   &  8.5    &   8.8   &  5.4    &  5.3    &  5.2    &   5.2    \\
1000  &  5.3    &   5.3   &  5.1    &  5.1    &  5.5    &  5.3    &  5.3    &   5.3    \\
\hline
\end{tabular}
\end{center}
\end{table*}

\begin{table*}
\caption{Properties of the generators: numbers of failed tests
for the batteries of tests SmallCrush, Crush, Bigcrush~\cite{TestU01},
and Diehard~\cite{TestU01}, and period lengths.
Testing was performed with
package TestU01 version TestU01-1.2.3.
For each battery of tests, we present three numbers: the number
of statistical tests with p-values outside the
interval $[10^{-3},1-10^{-3}]$, number of tests with p-values
outside the interval $[10^{-5},1-10^{-5}]$, and number of tests
with p-values outside the interval $[10^{-10},1-10^{-10}]$.
}
\begin{center}
\begin{tabular}{|l||c|c|c|c||c||c|c||}
\hline
Generator & SmallCrush & Diehard & Crush & Bigcrush & Period \\
\hline
MRG32k3a  & $ 0, 0, 0$ & $ 0, 0, 0$ & $ 0, 0, 0 $ & $ 0, 0, 0$ & $3.1\cdot 10^{57}$  \\
LFSR113   & $ 0, 0, 0$ & $ 1, 0, 0$ & $ 6, 6, 6 $ & $ 6, 6, 6$ & $1.0\cdot 10^{34}$ \\
MT19937   & $ 0, 0, 0$ & $ 0, 0, 0$ & $ 2, 2, 2 $ & $ 2, 2, 2$ & $4.3\cdot 10^{6001}$ \\
GM19SSE   & $ 0, 0, 0$ & $ 0, 0, 0$ & $ 0, 0, 0$ & $ 0, 0, 0$ & $2.7\cdot 10^{11}$ \\
GM31SSE   & $ 0, 0, 0$ & $ 0, 0, 0$ & $ 0, 0, 0$ & $ 0, 0, 0$ & $4.6\cdot 10^{18}$ \\
GM61SSE   & $ 0, 0, 0$ & $ 0, 0, 0$ & $ 0, 0, 0$ & $ 0, 0, 0$ & $5.3\cdot 10^{36}$ \\
\hline
\end{tabular}
\end{center}
\label{RNG-Properties}
\end{table*}

\section{Statistical properties and period lengths}

Applying hundreds of empirical tests (so-called batteries of tests)
allows to estimate statistical robustness of the generators.
As we mentioned above, here we considered only generators with
best statistical properties (see also Sec. 4.5.4 and Sec. 4.6.1 in~\cite{LEcuyerMain},
and Sec. 3 in~\cite{BarashShchur}).

Table~\ref{RNG-Properties} shows the results of applying the SmallCrush,
PseudoDiehard, Crush and Bigcrush batteries of tests taken from~\cite{TestU01},
to the generators of our library.
For each battery of tests, Table~\ref{RNG-Properties} shows
three numbers: the number
of statistical tests with p-values outside the
interval $[10^{-3},1-10^{-3}]$, number of tests with p-values
outside the interval $[10^{-5},1-10^{-5}]$, and number of tests
with p-values outside the interval $[10^{-10},1-10^{-10}]$.
The last column in Table~\ref{RNG-Properties} shows the RNG period
lengths.

Libraries SmallCrush, PseudoDiehard, Crush and Bigcrush contain 15, 126, 144
and 160 tests respectively.
We consider the statistical test ``failed'' if the p-value lies
outside the region $[10^{-3},1-10^{-3}]$.
Varying free initial conditions of the generators GM19, GM31 and GM61 
and applying statistical tests, one can observe occasionally a single failed test 
with p-value in the region $[10^{-5},10^{-3}]\cup [1-10^{-3},1-10^{-5}]$.
Since the number of applied tests is of the order of $10^3$, such a single failed test
can be explained simply by random statistical fluctuations, which do not characterize
the quality of random numbers.
As seen, all generators possess excellent statistical properties.
Table~\ref{RNG-Properties} shows that the generators, except LFSR113 and MT19937 
pass all statistical tests in the TestU01 library. Such generators can be recommended
for practical use.

As was mentioned in~\cite{TestU01Paper}, the generators MT19937 and LFSR113 fail tests
that try to find a linear structure in the bits of the output, because their bits have
a linear structure by construction.
Hence this particular point does not mean serious deficiencies of the generator MT19937.
MT19937 passes all other tests.
Namely, both of MT19937 and LFSR113 fail
the test \verb#scomp_LinearComp#, and also LFSR113 fails the test \verb#smarsa_MatrixRank#.

This work was supported by Russian Foundation for Basic Research.


\begin{thebibliography}{99}

\bibitem{cm} K.S.D.\ Beach, P.A.\ Lee, P.\ Monthoux, Phys. Rev.
Lett. {\bf 92} (2004) 026401; D.P.\ Landau and K.\ Binder, {\it A
Guide to Monte Carlo Simulations in Statistical Physics} (Cambridge
University Press, Cambridge, 2000); S.C.\ Pieper and R.B.\ Wiring,
Ann. Rev. Nucl. Part. Sci., {\bf 51} (2001) 53;A.\ L\"uchow, Ann. Rev.
Phys. Chem., {\bf 51} (2000) 501; A.R.\ Bizzarri, J. Phys.: Cond. Mat.
{\bf 16} (2004) R83.

\bibitem{Knuth} D.E.\ Knuth, {\em The art of Computer Programming},
Vol.\ 2, (Addison-Wesley, Cambridge, 1981).

\bibitem{MT} M.\ Matsumoto and T.\ Nishimura,
ACM Trans. on Mod. and Comp. Sim., {\bf 8} (1998) 3.

\bibitem{CombinedLCG}
P.\ L'Ecuyer, Oper. Res., {\bf 47} (1999) 159.

\bibitem{LFSR113}
P.\ L'Ecuyer, Math. of Comp., {\bf 68} (1999) 261.

\bibitem{BarashShchur} L.\ Barash, L.N.\ Shchur, Phys. Rev. E {\bf 73} (2006), 036701.

\bibitem{Pentium4}
\verb#http://www.intel.com/support/processors/pentium4#
\verb#/sb/CS-029967.htm#

\bibitem{AMD64}
\verb#http://support.amd.com/us/Processor_TechDocs#
\verb#/24592.pdf#

\bibitem{RS}
L.N.\ Shchur and T.A.\ Rostunov, JETP Lett., {\bf 76} (2002) 475.


\bibitem{AlgSite}
\verb#http://www.comphys.ru/rng_sse.htm#

\bibitem{Coveyou} R.R.\ Coveyou and R.D.\ MacPherson, J. ACM {\bf 14}
(1967) 100; G.\ Marsaglia, Proc. Nat. Acad. Sci. USA {\bf 61}
(1968) 25.

\bibitem{Kirk} S.\ Kirkpatrick and E.P.\ Stoll, 
J. Comp. Phys., {\bf 40} (1981) 517.

\bibitem{Golomb} S.W.\ Golomb, {\em Shift Register Sequences},
(Holden-Day, San Francisco, 1967).

\bibitem{SR1} A.M.\ Ferrenberg, D.P.\ Landau, Y.J.\ Wong, Phys.\ Rev.\
Lett., {\bf 69} (1992) 3382.

\bibitem{S99} L.N.\ Shchur, Comp.\ Phys.\ Comm., {\bf 121-122} (1999) 83.

\bibitem{Grass}
P.\ Grassberger, Phys. Lett., {\bf 181} (1993) 43.

\bibitem{RandomWalksTest} L.N.\ Shchur, J.R.\ Heringa, H.W.J.\
Bl\"ote, Physica A, {\bf 241} (1997) 579; L.N.\ Shchur, H.W.J.\ Bl\"ote,
Phys.\ Rev.\ E, {\bf 55} (1997) R4905.

\bibitem{SR3}
F.\ Schmid, N.B.\ Wilding, Int.J.Mod.Phys., C {\bf 6} (1995) 781.

\bibitem{LEcuyerMain}
P.\ L'Ecuyer,
Chapter 4 of the Handbook of Simulation, Jerry Banks Ed., Wiley, 1998,
pp. 93--137.

\bibitem{Arnold} V.I.\ Arnol'd, A.\ Avez, {\em Ergodic Problems of
Classical Mechanics}, (Benjamin, New York, 1968).

\bibitem{Grothe}
H.\ Grothe, Statistische Hefte, {\bf 28} (1987) 233.

\bibitem{Lecuyer90}
P.\ L'Ecuyer, Comm. of the ACM, {\bf 33(10)} (1990) 85.

\bibitem{NiederrBook}
H.~Niederreiter, in {\it Monte Carlo and Quasi-Monte Carlo Methods in
Scientific Computing}, ed. H. Niederreiter and P. J.-S. Shiue,
Lecture Notes in Statistics, (Springer-Verlag, 1995).

\bibitem{SIMD-MT}
M. Saito and M. Matsumoto, Monte Carlo and Quasi-Monte Carlo Methods 2006,
Springer, 2008, pp. 607--622.

\bibitem{MTInit} M.\ Matsumoto et. al.,
{\it Mersenne Twister with improved initialization},
\verb#http://www.math.sci.hiroshima-u.ac.jp/~m-mat/MT#
\verb#/MT2002/emt19937ar.html#

\bibitem{TestU01}
P.\ L'Ecuyer, R.\ Simard, {\it TestU01: A Software Library in ANSI C for
Empirical Testing of Random Number Generators (2002)}, Software
user's guide, \verb#http://www.iro.umontreal.ca/~simardr/testu01#
\verb#/tu01.html#

\bibitem{TestU01Paper}
P.\ L'Ecuyer, R.\ Simard, {\it TestU01: A C Library for Empirical Testing of
Random Number Generators},
ACM TOMS, {\bf 33(4)} (2007) article 22.

\end{thebibliography}
\end{document}